\documentclass[aps,pra,twocolumn,showpacs,preprintnumbers,amsmath,amssymb,superscriptaddress]{revtex4-1}

\usepackage{dsfont, color, braket}

\usepackage{graphicx}
\usepackage{bbold}
\usepackage{amsmath,amssymb,float,amsthm,mathrsfs}
\usepackage{bm}
\usepackage{epstopdf}

\begin{document}
\title{Charged oscillator quantum state generation with Rydberg atoms}
\author{Robin Stevenson}
\affiliation{School of Physics and Astronomy, The University of Nottingham, Nottingham, NG7 2RD, United Kingdom}
\author{Ji\v{r}\'{i} Min\'{a}\v{r}}
\affiliation{School of Physics and Astronomy, The University of Nottingham, Nottingham, NG7 2RD, United Kingdom}
\author{Sebastian Hofferberth}
\affiliation{5. Physikalisches Institut and Center for Integrated Quantum Science and Technology, Universit\"at Stuttgart, Pfaffenwaldring 57, 70569 Stuttgart, Germany}
\author{Igor Lesanovsky}
\affiliation{School of Physics and Astronomy, The University of Nottingham, Nottingham, NG7 2RD, United Kingdom}
\begin{abstract}
We explore the possibility of engineering quantum states of a charged mechanical oscillator by coupling it to a stream of atoms in superpositions of high-lying Rydberg states. Our scheme relies on the driving of a two-phonon resonance within the oscillator by coupling it to an atomic two-photon transition. This approach effectuates a controllable open system dynamics on the oscillator that permits the dissipative creation of squeezed and other non-classical states which are central to applications such as sensing and metrology or for studies of fundamental questions concerning the boundary between classical and quantum mechanical descriptions of macroscopic objects. We show that these features are robust to thermal noise arising from a coupling of the oscillator with the environment. Finally, we assess the feasibility of the scheme finding that the required coupling strengths are challenging to achieve with current state-of-the-art technology.
\end{abstract}

\maketitle

\section{Introduction}

The interface between different types of quantum systems has been the subject of much attention in the quest for complex quantum technologies \cite{Wallquist:09, Kolkowitz:12, Xiang:13,Ladd:10}. In order to combine advantages of various platforms, such as long coherence time, strong interactions or low-loss transport \cite{Aspelmeyer:13}, one has to be able to transfer quantum state between different systems. Alternatively, the interactions between two different quantum systems can be exploited to produce and probe quantum states \cite{Kuzmich:98, Lahaye:09}.

Mechanical systems in particular have seen rapid experimental progress. Nowadays, micro- and nano-mechanical oscillators can be cooled down to the quantum regime, where the quantised dynamics of the oscillator motion and controlled interaction with other quantum systems have become possible \cite{Kleckner:08, Poot:12, Aspelmeyer:13,Peterson:16}. An alternative to the typically used optomechanical interaction is to exploit electric forces to couple an atom to a charged oscillator \cite{Bariani:14, Rabl:09, Rabl:10, Seok:11,Sanz:15}. The strong dipole moment of atoms excited to high principal number  Rydberg states \cite{Saffman:09}, allows strong free-space interaction between single atoms and a charged oscillator, without the need for a mediating cavity. Atomic dipole - oscillator dipole coupling allows single atom cooling and the construction of complex superpostions of phononic Fock states \cite{Law:96}. Moreover, efficient coupling between Rydberg atoms and microwave cavities \cite{Hogan:12}, acceleration of flying atoms \cite{Lancuba:14} and creation of superpositions between different Rydberg states \cite{Facon:16} all constitute well established technologies. At the same time, results in the fabrication of micromechanical oscillators with resonance frequencies matching Rydberg transitions in atomic systems, and with high quality factor are promising, particularly using single-crystal diamonds \cite{Tao:13,Burek:12}. Additionally, these oscillators can be superconducting, and thus become chargeable on demand \cite{Bautze:14}. 

\begin{figure}[t!]
  	\includegraphics[width=8.6cm]{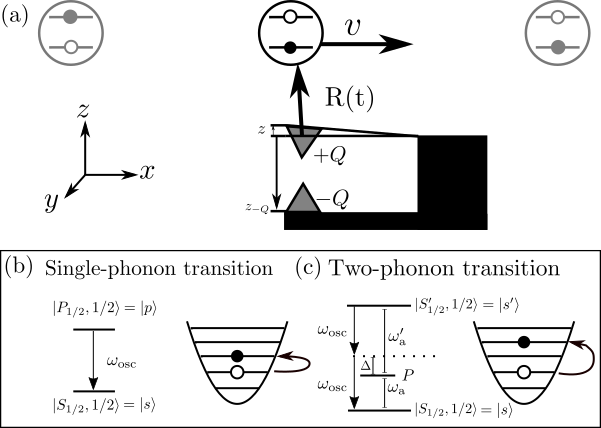}
  	\caption{(a) Setup of the system. Atoms pass one at a time above a micromechanical oscillator. An arm with charge $+Q$ oscillates vertically, while another arm with charge $-Q$ is fixed at position $z_{-Q}$. Atoms pass the oscillator at a rate $r$. See text for details. (b) In the single-phonon process one de-excitation of the atom excites a single-phonon transition in the oscillator. (c) In the two-phonon process a two-photon transition in the atom via an intermediate manifold excites a two-phonon transition in the oscillator.}
 	\label{fig:setup}
\end{figure}

In this paper we exploit the coupling between flying Rydberg atoms and a charged mechanical oscillator. We show that when the oscillator is driven at two-phonon resonance and if the coupling between the atoms and the oscillator is sufficiently strong, the system dynamics results in a non-classical state of the oscillator, whose nature can be tuned by a suitable choice of the initial atomic state. The desired oscillator states are obtained after the passage of only tens of atoms corresponding to the initial transient period of an effective dissipative dynamics. Specifically we show that under the strong coupling condition one can create a squeezed or Schr\"odinger cat states of the oscillator which are robust with respect to realistic thermal noise. These states are particularly useful for fundamental tests of quantum physics and decoherence processes \cite{Bose:99, Arndt:14}, quantum information and quantum simulation \cite{Lombardo:15}, metrology and sensing of small forces \cite{Gilchrist:04} or even for dark matter detection \cite{Bateman:15} or to probe quantum gravity inspired models \cite{Ghobadi:14}. While squeezed states of micromechanical oscillators have been produced \cite{Wollman:15,Pirkkalainen:15,Lecocq:15}, the creation of large and robust Schr\"odinger cat states of macroscopic mechanical oscillators is yet to be achieved. We perform a feasibility study and find that with current state-of-the-art technology it is challenging to access the strong coupling regime. To ultimately reach the desired coupling strengths may necessitate further developments, such as the use of collectively enhanced coupling through oscillator arrays or atomic ensembles.

The article is structured as follows. We introduce the system and describe its dynamics in Sec. \ref{sec:System}. We study the effect of thermal fluctuations and experimental contraints in Sec. \ref{sec:Exp}. Finally we comment on the implications of those constraints and discuss possible future directions in Sec. \ref{sec:Discussion}. 

\section{The system}
\label{sec:System}

The system under consideration is shown in Fig. \ref{fig:setup}(a). It consists of a stream of single Rydberg atoms coupled to a charge $Q$ at the tip of a micromechanical oscillator, which oscillates in the $z$ direction around the origin. We denote by $\hat z = z_\text{osc}(\hat a + \hat a^\dagger)$ the displacement operator of the oscillator, where $\hat a^\dagger$ and $\hat a$ are the bosonic phonon creation and annihilation operators, $z_{\rm osc} = \sqrt{\hbar/(2 m_\text{eff} \omega_\text{osc} )}$ the characteristic oscillator length, $m_\text{eff}$ the effective mass of the oscillator and $\omega_{\rm osc}$ is the mechanical oscillation frequency. The atoms move along a path $\mathbf{R}(t) = \left(X(t), Y(t)=0, Z(t)\right)$ such that only one atom is interacting with the oscillator at a time.

\subsection{Single atom dynamics}

In this article we consider two distinct situations: a \emph{single-phonon} and \emph{two-phonon} resonance (see Fig. \ref{fig:setup}). In the first case the atomic ground state $\ket{s}=|S_{1/2},1/2\rangle$, the excited state $\ket{p}=|P_{1/2},1/2\rangle$, $\omega_{\rm a}$ is the $\ket{s}-\ket{p}$ transition frequency and the interaction is described by the interaction Hamiltonian $\hat V = -\hat {\bm \mu} \cdot \mathbf{\hat E}[\mathbf{R}(t)]$, see Fig. \ref{fig:setup}(b). Here, $\hat {\bm \mu}$ is the atomic dipole of the $|s\rangle - |p\rangle$ transition and $\mathbf{\hat E}[\mathbf{R}(t)]$ is the electric field at the position $\mathbf{R}(t)$ created by the oscillator charge.
 In the latter case, the two-phonon oscillator transition couples to a two-photon transition between Rydberg levels $\ket{s}=|S_{1/2},1/2\rangle$ and $\ket{s'}=|S'_{1/2},1/2\rangle$, which are $S$ states with different principal quantum number, via an off-resonant manifold of $P$ states. We denote by $\omega_{\rm a}'$ the $P - S'$ transition frequency and by $\Delta = \omega_{\rm a}' - \omega_{\rm osc}$ the atom-oscillator detuning, which is assumed to be much larger than the energy separation of states within the $P$ manifold. The interaction Hamiltonian in this case reads $\hat V = -(\hat {\bm \mu}_2 + \hat {\bm \mu}'_2)\cdot \mathbf{\hat E}[\mathbf{R}(t)]$, where $\hat {\bm \mu}_2$ ($\hat {\bm \mu}'_2$) is the dipole moment of the $S - P$ ($P - S'$) transition (see Appendix \ref{app:interaction}).

\subsection{Single-phonon resonance}

The first scenario we are studying is that of a single-phonon resonance, where $\omega_{\rm osc} = \omega_{\rm a}$. Under the assumption of small oscillator displacement as compared to the distance between the oscillator and the flying Rydberg atom, $z \ll R$, where $R = |{\bf R}(t)|$, one can expand the electric field in powers of $\hat z$. Using the rotating wave approximation, the interaction picture Hamiltonian reads (see Appendix \ref{app:interaction})
\begin{align}
  \hat H_\text{I}(t) \approx \hbar \gamma(t) |s\rangle\langle p|\hat a^\dagger + \hbar \gamma^*(t) |p\rangle\langle s|\hat a, \label{eqn:Ham1}
\end{align}
where $\gamma(t) = \frac{Q \mu_0 z_\text{osc}}{4\pi\varepsilon_0 \hbar R^5 } \frac{( 3 Z^2 -R^2 )}{3}$ is the time dependent coupling strength \footnote{In principle, the form of the coupling between the atom and the oscillator depends on the geometry of a given mechanical mode and should be taken into account. For simplicity we assume that the atomic dipole couples to the center of mass motion of the mechanical oscillator as described in Appendix \ref{app:interaction}.}. For this resonant case, the time evolution can be solved exactly with the propagator $\hat U(t_f,t_i) = {\rm exp}(-i\hat H_\text{I}(t_f-t_i)/\hbar) = \sum_{n=0}^\infty \hat U_n(t_f,t_i)$, where
\begin{align}
\hat U_n(t_f,t_i) = 
  \begin{pmatrix}
   \cos\Theta_n & -i\sin\Theta_n \\
   -i \sin\Theta_n & \cos\Theta_n
  \end{pmatrix}. 
  \label{eq:U_n}
\end{align}
Here $n$ is the oscillator phonon occupation number, $\Theta_n = \sqrt{n+1}\, G$, $G = \int_{t_i}^{t_f} {\rm d}t \gamma(t)$ is the integrated coupling strength and (\ref{eq:U_n}) is written in the $\{\ket{p,n}, \ket{s,n+1}\}$ basis. This is a situation corresponding to the micromaser physics as described for example in Ref. \cite{Englert:02}.


The atoms are prepared identically and interact one at a time with the oscillator (see Fig. \ref{fig:setup}(a)) such that the evolution of the oscillator can be evaluated according to $\hat U(t_f,t_i)$ after the passage of each single atom. The initial state of each atom is assumed to be a superposition of the form
\begin{align}
 \ket{\psi}_{\rm a} = \alpha \ket{s} + \beta \ket{p}, \label{eqn:init}
\end{align}
with the amplitude $\beta = \sqrt{1 - |\alpha|^2} \, e^{i\theta}$. The state of the oscillator can be determined at an arbitrary time iteratively as follows: the state of the oscillator $\rho_\text{osc}^{(k)}$ after $k$ atoms have passed can be obtained by time-evolving the initial product state $\rho_\text{a}\otimes\rho_\text{osc}^{(k-1)}$ (where $\rho_\text{a} = |\psi\rangle_\text{a}\langle\psi|_{\rm a}$ is the initial state (\ref{eqn:init}) of the atom) with $\hat U$ and subsequently tracing out the atomic degrees of freedom
\begin{equation}
	\rho_\text{osc}^{(k)} = \text{Tr}_\text{a}[\hat U \rho_\text{a}\otimes\rho_\text{osc}^{(k-1)}\hat U^\dagger]. 
	\label{eq:evolution}
\end{equation}


The propagator $\hat U$ gives the exact evolution of the system as an atom travels past. However, it is useful to describe the dynamics of the oscillator in terms of an approximate master equation. We derive the master equation in the limit where the change in the oscillator state due to the interaction with a single atom is small such that $\dot \rho_{\rm osc} \approx r \Delta \rho_{\rm osc}^{(k)}$, where $\Delta \rho_\text{osc}^{(k)} = \rho_\text{osc}^{(k+1)} - \rho_\text{osc}^{(k)}$ and $r$ is the rate by which the atoms fly by the oscillator. The master equation approach has the advantage that it provides useful insights in the dynamics of the system without explicit exact solution. It also allows for adding directly the coupling to a thermal bath \cite{Englert:02}, as we shall discuss in detail in the case of the two-phonon resonance.
 
Next, assuming $\Theta_n \ll 1$, the propagator (\ref{eq:evolution}) can be expanded to second order in $\Theta_n$ which yields the effective open system dynamics
\begin{align}
\dot \rho_{\rm osc} \approx & -i r G \left[ \alpha \beta^* \hat a + \beta \alpha^* \hat a^\dagger, \rho_{\rm osc}\right] \nonumber\\ 
	& + r \Big( \mathcal D[\alpha G \hat a](\rho_{\rm osc}) + \mathcal  D[\beta G \hat a^\dagger](\rho_\text{osc}) \Big ),
	\label{eq:master single}
\end{align}
where $\mathcal D[\hat c](\rho) = \hat c \rho \hat c^\dagger - \frac{1}{2}\left(\hat c^\dagger \hat c \rho - \rho \hat c^\dagger \hat c\right)$ is the Lindblad dissipator (see Appendix \ref{app:master eq} for details).

\begin{figure*}[t!]
\includegraphics[width=17.2cm]{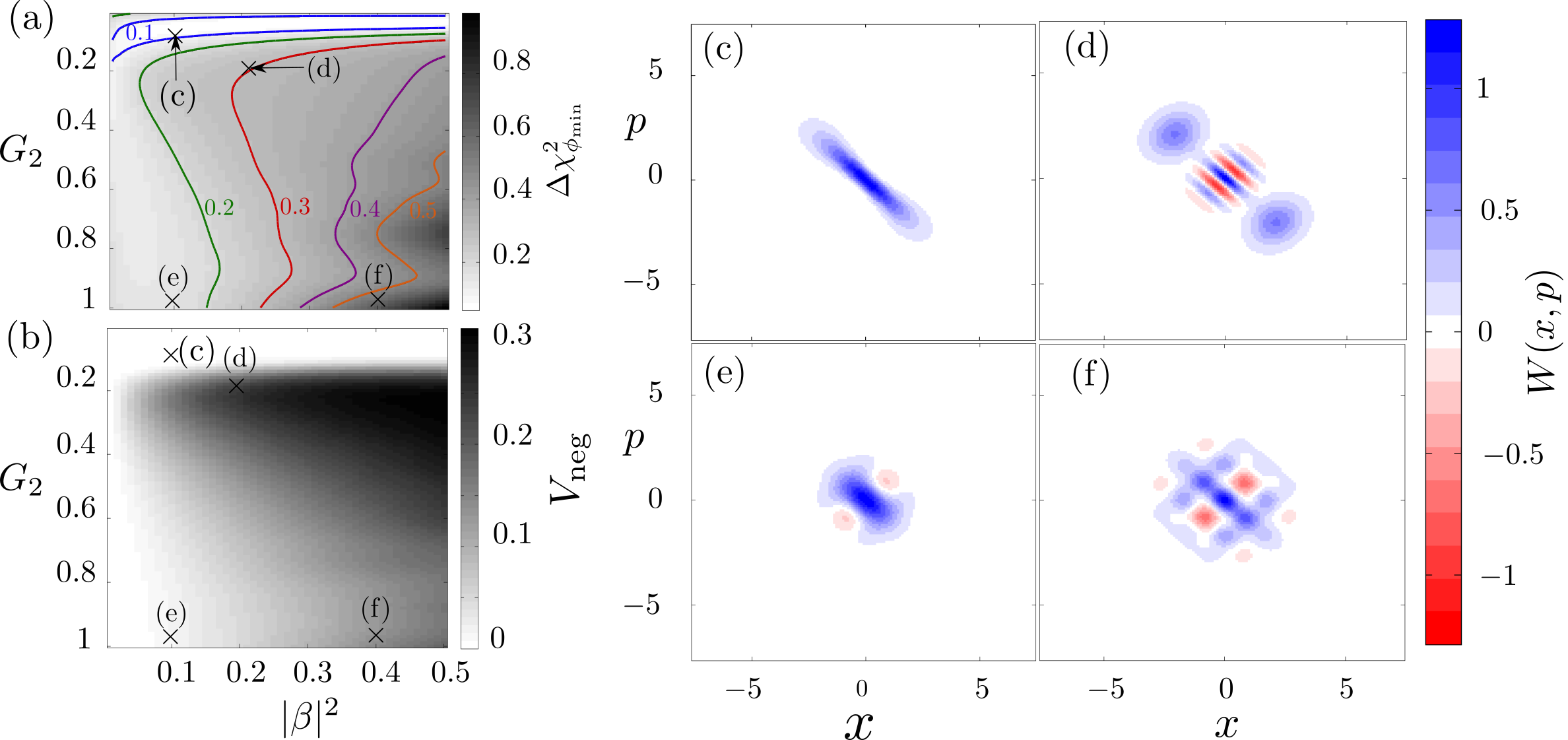}
  	\caption{(color online) (a) Minimum variance $\Delta \chi^2_{\phi_{\rm min}}$ and (b) negative volume of the Wigner function $V_\text{neg}$ after $k=30$ atoms have passed, as a function of atomic excited state population $|\beta|^2$ and two-phonon integrated coupling strength $G_2$. (c)-(f) Wigner function $W$ of the oscillator state for different parameter choices $(G_2,|\beta|^2)$: panel (c) $(0.06,0.1)$, panel (d) $(0.2,0.2)$, panel (e) $(1,0.1)$ and panel (f) $(1,0.4)$. \label{fig:Gbeta} }  	
\end{figure*}

\subsection{Two-phonon resonance}

In order to move beyond a displaced thermal state and achieve quantum states that are more complex we consider coherent two-phonon transitions of the oscillator that generate explicitly quantum effects.
On the atomic side we consider two possible coupling mechanisms: direct single-photon and intermediate states mediated two-photon transition between a pair of atomic levels. As both situations lead to equivalent form of the effective interaction Hamiltonian, we first focus only on the latter which we analyze in detail. We then invoke the former in Sec. \ref{sec:Exp} for the sake of quantitative comparison.

The situation for atomic two-photon transition via an intermediate manifold of states coupled to a two-phonon transition of the oscillator is depicted in Fig.\ref{fig:setup} (c). When the intermediate manifold of states is detuned far enough from resonance with a single phonon it remains unpopulated and can be eliminated from the dynamics leaving an effective two-level system.

In the following we consider the case of a two-phonon resonance with the initial atomic state $\ket{\psi}_{\rm a} = \alpha \ket{s} + \beta \ket{s'}$ as described in Fig.\ref{fig:setup} (c). On two-phonon resonance ($\omega_\text{a} + \omega_{\rm a}' = 2 \omega_\text{osc}$) the intermediate $P$ levels are adiabatically eliminated and the interaction between the atom and the oscillator is described by the interaction picture Hamiltonian
\begin{align}
  \hat H_\text{I,2}(t) \approx \hbar \gamma_2(t) |s\rangle\langle s'| (\hat a^\dagger)^2 + \hbar \gamma_2^*(t) |s'\rangle\langle s|\hat a^2 \label{eqn:Ham2}  
\end{align}
with $\gamma_2(t) =\left(\frac{Q z_\text{osc}}{ 4\pi\hbar\varepsilon_0 R^5}\right)^2\frac{\mu_0\mu'_0}{\Delta} \frac{1}{3}[R^2(R^2+3 Z^2)]$.

As in the single-phonon resonance case, the time evolution of the system can be solved exactly using the propagator $\hat U_2(t_f,t_i) = \exp(-i \hat H_\text{I,2} (t_f-t_i)/\hbar) = \sum_{n = 0}^\infty \hat U_{n,2}(t_f,t_i)$. Here

\begin{align}
  \hat U_{n,2}(t_f,t_i) =  \begin{pmatrix}
   \cos\Theta_{n,2} & -i \sin\Theta_{n,2} \\
   -i \sin\Theta_{n,2} & \cos\Theta_{n,2}
  \end{pmatrix}, 
  \label{eq:Ures2}
\end{align}
which is now written in the basis $\{|s',n\rangle,|s,n+2\rangle \}$ ,
 $\Theta_{n,2} =\sqrt{(n+1)(n+2)} \, G_2$  and $G_2 = \int_{t_i}^{t_f} {\rm d}t \gamma_2(t)$. Note that the evolution in the odd/even $n$ subspaces of the oscillator are independent of each other.

The two-phonon coupling between the atom and the oscillator is reminiscent of two-photon micromasers \cite{Gomes:99, Vidella:01, Ahmad:08}, and we show here that it allows the creation of squeezed states, as suggested by the form of the Hamiltonian (\ref{eqn:Ham2}) \cite{Nunnenkamp:10}. For the quantification of squeezing we introduce the standard quadrature observable
\begin{align}
\Delta \chi^2_\phi \equiv \braket{\hat \chi_\phi^2} - \braket{\hat \chi_\phi}^2,
\end{align}
where $\hat \chi_\phi = (\hat a e^{-i \phi} + \hat a^\dagger e^{i \phi})/\sqrt{2}$. The quadrature angles $\phi=0,\pi/2$ correspond to the $X$ and $P$ quadratures, and the state is squeezed along $\phi$ if $\Delta \chi^2_\phi < 1/2$. The squeezing of mechanical motion was in fact achieved in recent experiments \cite{Wollman:15,Pirkkalainen:15,Lecocq:15}. 
The manipulation of the oscillator state using Rydberg atoms at two-phonon resonance however goes beyond the squeezed state preparation and allows for creation of various other kinds of non-classical states. In order to quantify the non-classicality of the created states we use the negativity of the Wigner quasi-probability distribution $W(x,p) = \frac{1}{\pi \hbar} \int_{-\infty}^{\infty} dy \langle x+y|\rho_\text{osc}|x-y\rangle e^{i 2 p y /\hbar}$,  where $\langle \psi \ket{x} = \psi(x)$ is the spatial wavefunction of the oscillator \cite{Wigner:32}. The negative volume of the Wigner function then reads \cite{Kenfack:04}
\begin{align}
V_\text{neg} = \frac{1}{2}\left(\int dx\, dp\, |W(x,p)| - 1\right).\label{eqn:vNeg}
\end{align}

The exact evolution of the system can be solved by iteratively applying (\ref{eq:evolution}) where $\hat U$ is replaced by $\hat U_2$ and we take $\rho_\text{osc}^{(0)} = |0\rangle\langle 0|$. The resulting state depends on the number $k$ of atoms that pass by the oscillator. The exact value of $k$ is not particularly important, as long as the number of atoms is sufficient to reach the desired non-classical state. For the following calculations, we fix $k=30$, which fulfills this conditions for all considered states.

We now turn to numerical simulation of the exact evolution as described by eqs. (\ref{eqn:init}),(\ref{eq:evolution}) and (\ref{eq:Ures2}). The results of the simulation are summarized in Fig.~\ref{fig:Gbeta}. Fig.~\ref{fig:Gbeta}(a) shows the minimum variance $\Delta \chi^2_{\phi_{\rm min}}$ of the state of the oscillator as a function of the integrated coupling strength $G_2$ and the atomic excited state population $|\beta|^2$. The angle $\phi_{\rm min}$ minimizing $\Delta \chi^2_\phi$ depends only on the relative phase $\theta$ between the atomic states (see Appendix \ref{app:squeezing}). For $\theta=0$ used in the simulation, $\phi_{\rm min}=\pi/4$. The negative volume of the Wigner function $V_\text{neg}$ (\ref{eqn:vNeg}) is plotted in Fig.~\ref{fig:Gbeta}(b).

Finally, Fig.~\ref{fig:Gbeta}(c-f) show the Wigner function for specific values of $G_2$ and $|\beta|^2$ denoted by $\times$ in Fig.~\ref{fig:Gbeta}(a). Points (c) and (e) show examples of squeezed states for small $G_2$ and large $G_2$ respectively. Points (d) and (f) show examples of states with significant negative regions of the Wigner function. The state shown in Fig.~\ref{fig:Gbeta}(d) has the qualitative features of a cat state \cite{Dodonov:74}, which is of particular interest as it is used in metrology for small force sensing \cite{Gilchrist:04} and in fundamental test of quantum mechanics \cite{Arndt:14}.

\section{Thermal fluctuations and experimental considerations}
\label{sec:Exp}

We now investigate how robust the production of these quantum states is in the presence of thermal fluctuations. Combining the master equation for the interaction with the passing atoms, derived analogously to the single phonon case (see Appendix \ref{app:master eq}), with the thermal processes gives
\begin{equation}
 \dot \rho_\text{osc} \approx \mathcal L_\text{a}[\rho_\text{osc}] + \mathcal L_\text{th}[\rho_\text{osc}],
 \label{eq:master two}
\end{equation}
where the atomic part is
\begin{align}
\mathcal L_\text{a}[\rho_\text{osc}] &= r \Big( -i G_2 \left[ \alpha \beta^* \hat a^2 + \beta \alpha^* (\hat a^\dagger )^2, \rho_\text{osc}\right] \nonumber\\
			& + \mathcal D[\alpha G_2 \hat a^2](\rho_\text{osc}) + \mathcal  D[\beta G_2 (\hat a^\dagger)^2](\rho_\text{osc}) \Big) \label{eqn:isolated}
\end{align}
and the thermal part is
\begin{align}
\mathcal L_\text{th} [\rho_\text{osc}] =  \Gamma_\text{m} (\bar n_{\rm th} + 1) \mathcal D[\hat a](\rho_\text{osc}) 
				+ \Gamma_\text{m}\bar n_{\rm th} \mathcal D[\hat a^\dagger](\rho_\text{osc}). \label{eqn:therm}
\end{align}
Here $\Gamma_{\rm m}$ is the coupling of the oscillator to the thermal bath and $\bar n_{\rm th} = \frac{1}{e^{\hbar \omega_{\rm osc}/k_\text{B} T}-1}$ is the mean phonon number of the bath at temperature $T$.

\begin{figure}[t!]
  	\includegraphics[width=8.6cm]{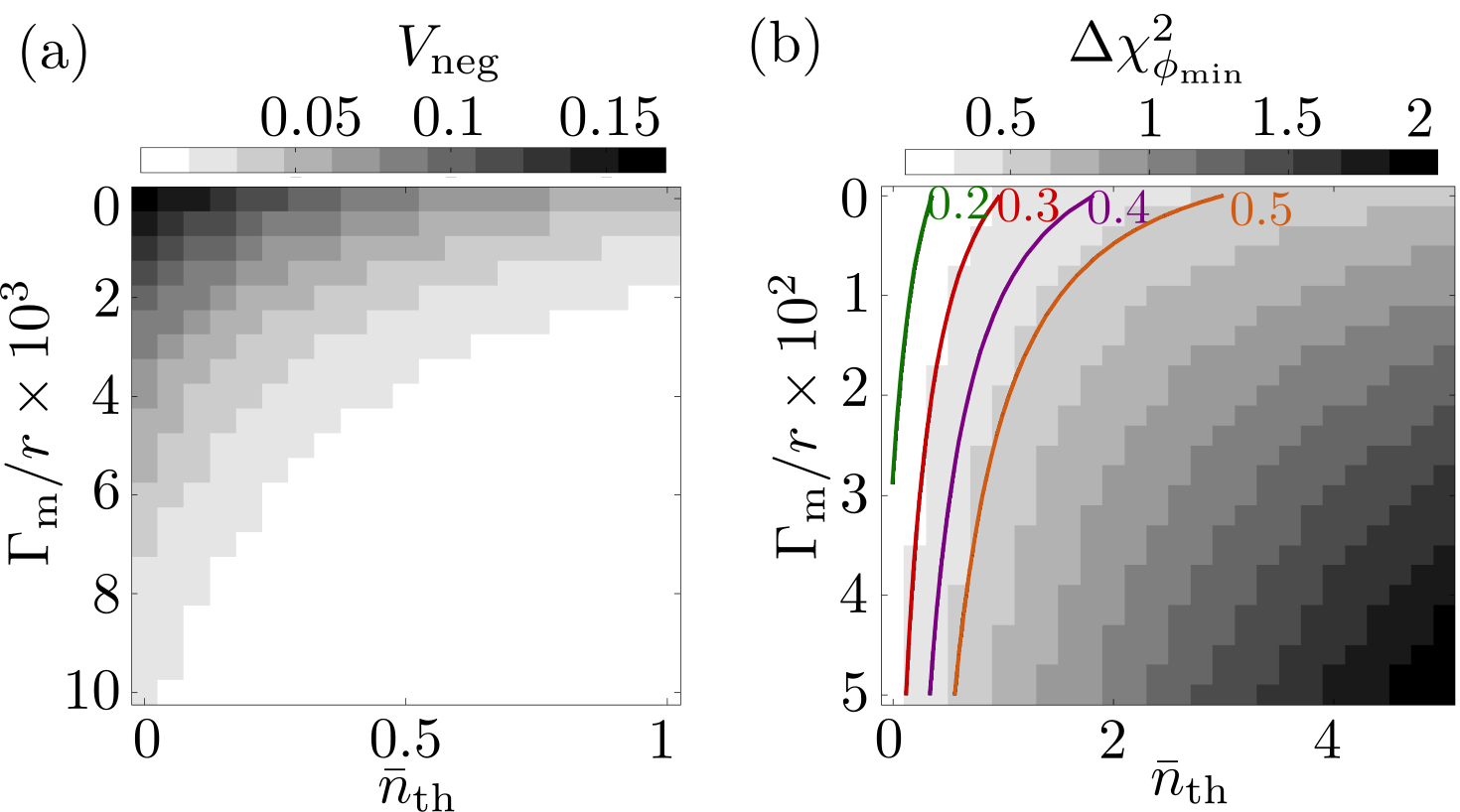}
  	\caption{(color online) (a) Negative volume $V_{\rm neg}$ of the Wigner function of the charged oscillator state and (b) minimum variance $\Delta \chi^2_{\phi_{\rm min}}$  after 30 atoms have passed as a function of relative thermal coupling strength $\Gamma_\text{m}/r$ and thermal bath occupation $\bar n_{\rm th}$. Parameters $(G_2,|\beta|^2)$ used: (a) $(0.2,0.2)$, (b) $(0.06,0.1)$.}
  	\label{fig:kNVn}
\end{figure}

To demonstrate how the coupling to the thermal bath deteriorates the oscillator quantum states, we solve the master equation (\ref{eq:master two}) numerically for a total time corresponding to the passage of 30 atoms and initial thermal state with $\bar{n}_{\rm th}$ \footnote{In practice, it might be difficult to control deterministically the exact number of the atoms interacting with the oscillator. We have verified, that varying the number of passing atoms from 25 to 35 results in only small changes (order of \%) of $V_{\rm neg}$.} . In Fig.~\ref{fig:kNVn}(a) we plot the negative volume of the Wigner function $V_{\rm neg}$ as a function of thermal coupling $\Gamma_\text{m}$ and mean thermal occupation number $\bar{n}_\text{th}$. The used parameters $G_2 = 0.2$, $|\beta|^2 = 0.2$ correspond to the cat state in Fig.~\ref{fig:Gbeta}(d). Fig.~\ref{fig:kNVn}(b) shows the minimum variance $\Delta \chi^2_{\phi_{\rm min}}$ for parameters corresponding to the squeezed state in Fig.~\ref{fig:Gbeta}(c).

It follows from Fig.~\ref{fig:Gbeta}(a,b) and Fig.~\ref{fig:kNVn}(a,b) that in order to create a non-classical state one requires $G_2 \sim 0.1$ and 
the atom passage rate $r$ should be maximized while minimizing $\Gamma_{\rm m}$ and $\bar n_{\rm th}$. With the help of specific examples, we demonstrate the performance of the scheme below.

Considering $\Gamma_\text{m} = 2\pi \times 500$ Hz and the state-of-the-art temperature $T=10$ mK corresponding to $\bar{n}_{\rm th} = 0.1$, we find for the cat state of Fig~\ref{fig:kNVn}(a) that $V_\text{neg} =0.25 V_\text{neg,0}$. Here $V_{\rm neg,0}$ denotes the value of $V_{\rm neg}$ for the system not coupled to a thermal bath ($V_{\rm neg,0}=0.24$ for the parameters used in Fig.~\ref{fig:kNVn}(a)). Similarly, using (\ref{eq:master two}) with the parameters from Fig.~\ref{fig:kNVn}(b), we find that for squeezing to be achieved one needs $\bar n_{\rm th} \lesssim 6$ corresponding to $T \lesssim 150$ mK. 

Next, in order to assess what couplings can be achieved in a realistic experiment, we consider the following parameters: $^{133}$Cs Rydberg atoms with a transition between $n = 100$ and $n=101$ which are separated by $\omega_\text{a} + \omega'_\text{a} \approx 2\pi \times 6$ GHz \cite{Lorenzen:83} corresponding to an oscillator resonant frequency $\omega_\text{osc} =  2\pi \times3$ GHz, which are achievable e.g. with clamped mechanical beams \cite{Huang_2004} (although with smaller quality factors than assumed in this work) or single-crystal diamond nanobars \cite{Burek:12}. The detuning between the oscillator frequency and the $P-S'$ transition frequency is $\Delta = \omega'_\text{a} - \omega_\text{osc} \approx 2\pi \times 300$ MHz, while the splitting  $P_{3/2} - P_{1/2} < 20$ MHz \cite{Lorenzen:83}. We take the oscillator characteristic length $z_\text{osc} = 10^{-13}$ m, the charge on the tip of the oscillator $Q = 200 e$ (compatible e.g. with $\sim$ aF capacitances of micron size electromechanical resonators operated with $\sim$ V voltages \cite{Yao_2000,LaHaye_2004}) and thermal bath coupling strength $\Gamma_\text{m}  = 2\pi \times 500$ Hz (corresponding to a quality factor $\mathcal{Q}=6\times10^6$ of the oscillator \cite{Burek:12}). For $n\approx 100$ Rydberg states the atomic size is $\approx 10^4 a_0 \approx 1\;\mu$m, and the corresponding dipole moments are $\mu_0 \approx \mu'_0 \approx 10^4 e a_0$ ($a_0$ is the Bohr radius). For the atomic motion, we consider a simple linear trajectory $\mathbf{R}(t)=(vt,0,Z_0)$ with $t$ going from $-\infty$ to $\infty$, where we neglect any deflection of the atom's path due to the interaction with the oscillator (the static monopole part of the field resulting from the charge Q can always be compensated by additional static charges; see Appendix \ref{app:interaction} for details of the interaction). Assuming the atom-cantilever distance to be $Z_0 = 5\;\mu$m we choose the atomic speed $v = 10$ m/s and the rate of atoms $r = 10^5$ atoms per second, giving the separation between successive atoms of $100\;\mu$m and the interaction time of couple of $\mu$s. This guarantees, to a good approximation, that only one atom is interacting with the oscillator at a time and that one can neglect the decay of the Rydberg states which have lifetime of 100 $\mu$s \cite{Feng:2009, Gallagher_1994}. We then obtain for the integrated coupling strength $G_2 = \left(\frac{Q z_\text{osc}}{4\pi\hbar\varepsilon_0}\right)^2\frac{\mu_0\mu'_0}{\Delta} \frac{21 \pi}{48 v Z_0^5} \approx 10^{-5}$. 

We now turn our attention to the direct single-photon two-phonon resonance provided by the atomic dipole - oscillator quadrupole coupling as we show in Appendix \ref{app:interaction}. Here, an analogous derivation leads to the integrated coupling strength $G_{2,\rm quad} = \frac{Q \mu_0 z_\text{osc}^2}{4 \pi \epsilon_0 \hbar} \frac{2}{3 v Z_0^3} \approx 10^{-9}$ which is smaller by orders of magnitude compared to $G_2$ in the two-photon two-phonon resonance scheme. 
~\\

\section{Discussion and Outlook}
\label{sec:Discussion}

We have explored a method of creating squeezed and non-classical states of a charged macroscopic mechanical oscillator. Such on-demand quantum state preparation constitutes a basic element of the mechanical oscillators state manipulation toolbox using atoms. Specifically, the squeezed and Schr\"odinger cat states that can be in principle generated might find applications as probes of decoherence processes of macroscopic bodies, in quantum information processing or in sensing and metrology. The values of the estimated couplings that are achievable with current state-of-the-art technology and typical parameter regimes turn out to be too small to be of a practical use. Further improvement might be sought e.g. by increasing the charge of the oscillator or by more suitable choice of the employed Rydberg states which would increase the dipole moment and decrease the two-photon detuning $\Delta$. Another possibility is to exploit the enhancement of the coupling when considering an ensemble of atoms coupled to an array of oscillators which we leave for future investigations.

\emph{Remark:} After finishing our manuscript, we became aware of a related work \cite{Adbi:16}.

\section{Acknowledgments}

J. M. would like to thank M. Marcuzzi and A. Armour for useful discussions. R. S. would like to thank A. Kouzelis for his comments. We thank W. Li for his help and the feedback on the manuscript. The research leading to these results has received funding from the European Research Council under the European Union's Seventh Framework Programme (FP/2007-2013) / ERC Grant Agreement n. 335266 (ESCQUMA). S.H. was supported by the German Research Foundation (DFG) through Emmy-Noether-grant HO 4787/1-1 and within SFB/TRR21.



\appendix

\begin{widetext}

\section{Atom-oscillator interaction}
\label{app:interaction}

The interaction Hamiltonian between the dipole moment $\hat{\bm{\mu}} = \{\hat{\mu}_{\text x}, \hat{\mu}_{\text y}, \hat{\mu}_{\text z}\} = \mu_0\{\hat M_{\text x},\hat M_{\text y},\hat M_{\text z}\}$ of an atom at position $\mathbf{R}$ and the electric field $\hat{\mathbf{E}}(\mathbf{R})$ created by a charge at position $\hat z$ can be expressed as a power series in $\hat z$ 

\begin{align}
  \hat V&=-\hat{\bm \mu}\cdot \mathbf{\hat E}(\mathbf{R}) \label{eqn:muE}\\
  &=-\frac{Q}{4\pi\varepsilon_0}
  		\frac{\hat \mu_{\text x} X+\hat \mu_{\text y}Y+\hat \mu_{\text z}(Z-\hat z) }
  					{[(X^2+Y^2+(Z-\hat z)^2]^\frac{3}{2}}
  					\\
  &\approx		 -\frac{Q \hat{\bm{\mu}}\cdot \mathbf{R}}{4\pi\varepsilon_0 R^3}
  				-\frac{Q\mu_0}{4\pi\varepsilon_0 R^5}
  				\Big[\hat M_{\text z} (3 Z^2-R^2)+
  					3 Z(\hat M_\text{x} X + \hat M_\text{y} Y) \Big]\hat z + O(\hat z^2) \label{eqn:DipExp} \\
  	&\equiv		 \hbar \gamma_0(\mathbf{R}) - \hbar \sum_{j = x,y,z} \left(\gamma_{j}(\mathbf{R}) \hat M_j (\hat a + \hat a^\dagger)\right) + O((\hat a + \hat a^\dagger)^2)  			 	
\end{align}
with $R = |\mathbf{R}|$ and the last line introduces notation for the coupling strengths $\gamma_{j}$ that are used in the following. 
The first term in eq. (\ref{eqn:DipExp}) corresponds to a Coulomb interaction, which can be cancelled by additional static charges with opposite sign (see also Fig. \ref{fig:setup}(a)) and thus we omit it in the following. The matrices $\hat M_\text{x,y,z}$ depend on the specific atomic transitions that couple to the electric field of the oscillator. We compute the matrix elements using the standard angular momentum theory as \cite{Messiah_1999, Friedrich_2006}
\begin{equation}
	\left(M_\alpha \right)^{L'_{J'},m_J'}_{L_{J},m_J} = \bra{L'_{J'},m_J'} \hat{\chi}_\alpha \ket{L_{J},m_J},
\end{equation}
where $\alpha=x,y,z$, $L$ is the electron angular momentum, $J$ the total angular momentum and $m_J$ the projection of the total angular momentum on the $z$ axis. The operators $\hat{\chi}$ are given by the relations $\hat{\chi}_{\pm 1} = \mp \frac{1}{\sqrt{2}}\left(\hat{\chi}_x \pm i \hat{\chi}_y\right)$ and $\hat{\chi}_0 = \hat{\chi}_z$. When expressed in the coordinate basis, they are simply rescaled spherical harmonics $\braket{\hat{\chi}_\pm} = \sqrt{\frac{4 \pi}{3}} Y_{1,\pm 1}(\theta,\phi)$, $\braket{\hat{\chi}_0} = \sqrt{\frac{4 \pi}{3}} Y_{1,0}(\theta,\phi)$. The dipole matrix elements are then obtained with the help of the relation
\begin{eqnarray}
	&& \bra{L'_{J'},m_J'} \hat{\chi}_q \ket{L_{J},m_J} = \nonumber \\ 
	&& = (-1)^{J'-m_J'} 
	\left( \begin{matrix}
		J' & 1 & J \\
		-m_J' & q & m_J
	\end{matrix} \right)
	(-1)^{J+S'+1}\sqrt{(2 J'+1)(2 J+1)}
	\left\{ \begin{matrix}
		L' & 1 & L \\
		J & S' & J'
	\end{matrix} \right\}
	\sqrt{(2 L'+1)(2 L+1)}
	\left( \begin{matrix}
		L' & 1 & L \\
		0 & 0 & 0
	\end{matrix} \right),
\end{eqnarray}
where $q=-1,0,1$, $S' = J'-L'$ is the total spin and $\left( \begin{matrix}
		\cdot & \cdot & \cdot \\
		\cdot & \cdot & \cdot
	\end{matrix} \right)$, $\left\{ \begin{matrix}
		\cdot & \cdot & \cdot \\
		\cdot & \cdot & \cdot
	\end{matrix} \right\}$ are the Wigner $3j$ and $6j$ symbols respectively. Note that we have absorbed the radial part of the dipole transition elements into the dipole moment amplitude $\mu_0$.

\subsection{Single-phonon transition}
\label{app:single-phonon}

For a single-phonon transition we consider resonant transitions the $S$ and $P$ manifolds of an atom within the same principal quantum number, as shown in figure \ref{fig:LevelStructure}(a). The transition matrices in the $|L_J,m_J\rangle = \{|S_{1/2}, -1/2\rangle,|P_{1/2}, -1/2\rangle,|S_{1/2}, 1/2\rangle,|P_{1/2}, 1/2\rangle\}$ basis read

\begin{align}
  \hat M_{\text x}  = -\frac{1}{3}\begin{pmatrix}
   	0 	& 0		& 0 	& 1 \\
	0 	& 0		& 1 	& 0 \\
	0	& 1 	& 0 	& 0 \\
	1	& 0 	& 0 	& 0 \\   
     \end{pmatrix}, 
     \hat M_{\text y}  =  -\frac{1}{3}\begin{pmatrix}
   	0 	& 0 	& 0 	& i \\
	0 	& 0		& i 	& 0 \\
	0	& -i 	& 0 	& 0 \\
	-i	& 0 	& 0 	& 0 \\   
     \end{pmatrix}, 
     \hat M_{\text z}  =  -\frac{1}{3}\begin{pmatrix}
   	0 	& -1 	& 0 	& 0 \\
	-1 	& 0		& 0 	& 0 \\
	0	& 0 	& 0 	& 1 \\
	0	& 0 	& 1 	& 0 \\   
     \end{pmatrix}, 
\end{align}

\begin{center}
	\begin{figure}[h!]
  	\includegraphics[width=16cm]{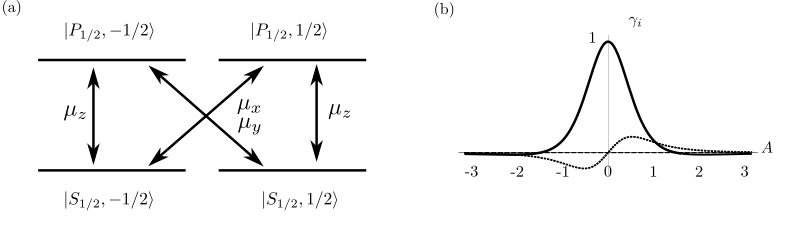}
  	\caption{(a) Level scheme and transitions for a four-level manifold. (b) Coupling strength $\gamma_{i}$ for $i=x$, $y$ and $z$ in short-dashed, long-dashed and solid lines respectively for an atom at position $\mathbf{R} = \{A Z_0, 0 ,Z_0\}$. The coupling strength has been scaled to the maximum of $\gamma_{z}$.}
 	\label{fig:LevelStructure}
 	\end{figure}
\end{center}

We calculate the coupling strengths for a position $\mathbf{R}(A) = \{A Z_0, 0, Z_0\}$. We find that $\gamma_{\text y}(\mathbf{R}(A)) = 0$ and that $\gamma_\text{z}(\mathbf{R}(A))/\gamma_{\text x}(\mathbf{R}(A)) = \frac{2 - A^2}{A}$ . Note that the last ratio is independent of $Z_0$. Fig.~\ref{fig:LevelStructure}(b) shows the dipole coupling strengths $\gamma_{j}(\mathbf{R}(A))$ as a function of the scaled coordinate $A$, where the coupling strengths have been normalized to the maximum value of $\gamma_\text{z}$. Since $\gamma_{\text z} \gg \gamma_{\text x}$ we neglect $\gamma_{\text x}$. This simplifies the description so that one can use only two of the four levels and we choose $|s\rangle \equiv |S_{1/2}, 1/2\rangle$ and $|p\rangle \equiv |P_{1/2}, 1/2\rangle$.

With this two-level system the atom-oscillator Hamiltonian can be written as $\hat H= \hat H_0 + \hat V$, where $\hat H_0 = \hbar \omega_\text{osc}\hat a^\dagger \hat a + \hbar \omega_\text{a} \hat \sigma^\text{z}$ and
\begin{align}
  \hat V &=  \frac{Q\mu_0 z_\text{osc}}{4\pi\varepsilon_0 R^5}
  				\frac{1}{3}\Big[\hat \sigma^{\text x} (3 Z^2-R^2)\Big](\hat a + \hat a^\dagger) + O([z_\text{osc}(\hat a + \hat a^\dagger)]^2).
  			 	\label{eqn:muE2}
\end{align}
where $\hat \sigma^\text{z} = |p\rangle \langle p| - |s\rangle\langle s|$, $\hat \sigma^\text{x} = |p\rangle \langle s| + |s\rangle\langle p|$ and we have used $\hat z = z_\text{osc}(\hat a + \hat a^\dagger)$. When $\omega_\text{osc} =\omega_\text{a}$ the $|s\rangle$-$|p\rangle$ transition of the atom is resonant with the one-phonon transition of the oscillator and the interaction picture Hamiltonian $\hat H_\text{I} = \text{exp}(-i \hat H_0 t/\hbar) \hat H \text{exp}(i \hat H_0 t/\hbar)$ reads

\begin{align}
  \hat H_\text{I}&=  \frac{1}{3} \frac{Q \mu_0 z_\text{osc}}{4\pi\varepsilon_0 R^5}\Big[				
  					 (|s\rangle\langle p|e^{-i \omega_\text{osc} t}
  					   + |p\rangle\langle s|e^{i \omega_\text{osc} t} )
  					   							( 3 Z^2 -R^2)
  					 \Big]  							
  						(   \hat a e^{-i \omega_\text{osc} t} 
  						  + \hat a^\dagger  e^{i \omega_\text{osc} t} )+  F \nonumber \\
			&=  \frac{1}{3} \frac{Q \mu_0 z_\text{osc}}{4\pi\varepsilon_0 R^5 }\Big[				
  				|s\rangle\langle p| \hat a^\dagger  
  					   	 ( 3 Z^2 -R^2 ) + |p\rangle\langle s|\hat a
  				   			(3 Z^2 - R^2 )\Big]+ F',\nonumber\\
  				   			&\approx \hbar \gamma(t) |s\rangle\langle p| \hat a^\dagger + \text{H. c.}
\end{align}
where $F$ and $F'$ contain only terms oscillating at $\omega_\text{osc}$ or higher frequency, which can be neglected through the rotating wave approximation and we have introduced the single-phonon coupling strength $\gamma(t) = \frac{Q \mu_0 z_\text{osc}}{4\pi\varepsilon_0 \hbar R^5 } \frac{( 3 Z^2 -R^2 )}{3} $.

\subsection{Two-phonon resonance}
\label{app:two-phonon}

Here we consider a situation where a two-photon atomic transition between different principal quantum number S-states $|s\rangle = |S_{1/2}, 1/2 \rangle$ and $|s'\rangle = |S'_{1/2},1/2\rangle$ couples to a two-phonon oscillator transition. The two-phonon transition is mediated by an off-resonant coupling to the $P_{1/2}$ and $P_{3/2}$ manifolds. We denote by $\omega_\frac{1}{2}$, ($\omega_{\frac{3}{2}}$) the $P_{1/2}-S'$ ($P_{3/2}-S'$) transition frequencies and by $\Delta_j = \omega_j - \omega_{\rm osc}$, $j=\frac{1}{2},\frac{3}{2}$ the respective detunings. In what follows, we refer to both manifolds combined as to the $P$ manifold. Formally, the $S-P$ ($P-S'$) transitions are described by a dipole moment operator $\hat{ \bm \mu}_2$ ($\hat{ \bm \mu}'_2$) with magnitude $\mu_0$ ($\mu_0'$) respectively. The atom-oscillator interaction is given by the sum $\hat V(\hat {\bm \mu}_2) + \hat V(\hat {\bm \mu}'_2)$, where $\hat V$ is given by (\ref{eqn:DipExp}), and the transition matrices $\hat M_j$ ($\hat M'_j$), $j=x,y,z$ for the $S-P$ ($P-S'$) transitions now read

\begin{align}
\hat M_\text{x} &=   \ket{s} \left(-\frac{1}{3}\langle P_{1/2},-1/2| + \frac{1}{3\sqrt{2}}\langle P_{3/2},-1/2| -\frac{1}{\sqrt{6}} \langle P_{3/2},3/2|\right) + \text{H.c} \nonumber\\
\hat M_\text{y} &=   i \ket{s} \left(\frac{1}{3}\langle P_{1/2},-1/2| - \frac{1}{3\sqrt{2}}\langle P_{3/2},-1/2| -\frac{1}{\sqrt{6}} \langle P_{3/2},3/2|\right) + \text{H.c} \nonumber\\
\hat M_\text{z} &= \ket{s} \left(-\frac{1}{3}\langle P_{1/2},1/2| + \frac{\sqrt 2}{3} \langle P_{3/2},1/2|\right) + \text{H.c} \nonumber\\
\hat M'_\text{x} &=   \ket{s'} \left(-\frac{1}{3}\langle P_{1/2},-1/2| + \frac{1}{3\sqrt{2}}\langle P_{3/2},-1/2| -\frac{1}{\sqrt{6}} \langle P_{3/2},3/2|\right) + \text{H.c} \nonumber\\
\hat M'_\text{y} &=   i \ket{s'} \left(\frac{1}{3}\langle P_{1/2},-1/2| - \frac{1}{3\sqrt{2}}\langle P_{3/2},-1/2| -\frac{1}{\sqrt{6}} \langle P_{3/2},3/2|\right) + \text{H.c} \nonumber\\
\hat M'_\text{z} &= \ket{s'} \left(-\frac{1}{3}\langle P_{1/2},1/2| + \frac{\sqrt 2}{3} \langle P_{3/2},1/2|\right) + \text{H.c}
\end{align}
\noindent To first order in $\hat z$, the total Hamiltonian in the atomic basis $\{|s\rangle, |s'\rangle, |P_{1/2},-1/2\rangle,|P_{1/2},1/2\rangle,P_{3/2},-1/2\rangle,$
$|P_{3/2},1/2\rangle,|P_{3/2},3/2\rangle\}$ reads
\begin{align}
  \hat H&= \hbar \begin{pmatrix}
   	0 	& 0					& 0					& 0					& 0&0&0  \\
	0	&2 \omega_\text{osc}& 0					& 0					& 0&0&0\\
	0	& 0 				& \omega_\text{osc} + \Delta_\frac{1}{2}	& 0 				& 0&0&0 \\
	0	& 0 				& 0 				& \omega_\text{osc} + \Delta_\frac{1}{2} 	& 0&0&0 \\
	0	& 0 				& 0 				& 0 				& \omega_\text{osc} + \Delta_\frac{3}{2}&0&0 \\  
	0&0	& 0 				& 0 				& 0 				& \omega_\text{osc} + \Delta_\frac{3}{2} &0\\ 
	0&0&0	& 0 				& 0 				& 0 				& \omega_\text{osc} + \Delta_\frac{3}{2} \\  
     \end{pmatrix}\nonumber\\
     &\hspace{10mm}     +\hbar \omega_\text{osc} \hat a^\dagger\hat a
    -\hbar\begin{pmatrix}
   	0 				& 0					&-\gamma_-		& -\gamma_\text{z}	&\frac{1}{\sqrt 2}\gamma_-		& \sqrt 2 \gamma_\text{z}& -\sqrt{\frac{3}{2}}\gamma_+  \\
	0				& 0					& -\gamma'_-		& -\gamma'_\text{z}& \frac{1}{\sqrt 2}\gamma'_-		& \sqrt 2 \gamma'_\text{z}	& -\sqrt{\frac{3}{2}}\gamma'_+\\
	-\gamma_+		& -\gamma'_+ 		& 0		& 0 				& 0 &0&0\\
	-\gamma_\text{z}	& -\gamma'_\text{z} 	& 0 			& 0			& 0 &0&0\\
		\frac{1}{\sqrt 2}\gamma_+		& \frac{1}{\sqrt 2}\gamma'_+ 		& 0		& 0 				& 0 &0&0\\
	\sqrt 2 \gamma_\text{z}	& \sqrt 2 \gamma'_\text{z} 	& 0 			& 0			& 0 &0&0\\
	-\sqrt{\frac{3}{2}}\gamma_-		& -\sqrt{\frac{3}{2}}\gamma'_- 		& 0 			& 0 				&0&0& 0\\   
     \end{pmatrix} (\hat a + \hat a^\dagger), 
  			 	\label{eqn:V2}
\end{align}


\noindent with $\hat z = z_\text{osc} (\hat a + \hat a^\dagger)$, 
and the atom-oscillator coupling strengths
$\gamma_\pm = \frac{Q\mu_0 z_\text{osc} }{4\pi\varepsilon_0 \hbar R^5} Z(X \pm iY)$, $\gamma_z = \frac{Q\mu_0 z_\text{osc} }{4\pi\varepsilon_0 \hbar R^5} \frac{3 Z^2 - R^2}{3}$ and similarly for $\gamma'_\pm,\gamma'_z$, where $\mu_0$ is replaced by $\mu_0'$. Taking the rotating wave approximation, the interaction picture Hamiltonian is
\begin{align}
  \hat H_\text{I}&= 
    -\hbar \begin{pmatrix}
   	0 				& 0					&-\gamma_- {\hat a}^\dag		& -\gamma_\text{z} {\hat a}^\dag	&\frac{1}{\sqrt 2}\gamma_- {\hat a}^\dag		& \sqrt 2 \gamma_\text{z} {\hat a}^\dag & -\sqrt{\frac{3}{2}}\gamma_+ {\hat a}^\dag  \\
	0				& 0					& -\gamma'_- {\hat a}		& -\gamma'_\text{z} {\hat a}& \frac{1}{\sqrt 2}\gamma'_- {\hat a}		& \sqrt 2 \gamma'_\text{z} {\hat a} & -\sqrt{\frac{3}{2}}\gamma'_+ {\hat a}\\
	-\gamma_+ {\hat a}		& -\gamma'_+ {\hat a}^\dag		& \Delta_\frac{1}{2}		& 0 				& 0 &0&0\\
	-\gamma_\text{z} {\hat a}	& -\gamma'_\text{z} {\hat a}^\dag	& 0 			& \Delta_\frac{1}{2}			& 0 &0&0\\
		\frac{1}{\sqrt 2}\gamma_+ {\hat a}		& \frac{1}{\sqrt 2}\gamma'_+ {\hat a}^\dag		& 0		& 0 				& \Delta_\frac{3}{2} &0&0\\
	\sqrt 2 \gamma_\text{z} {\hat a}	& \sqrt 2 \gamma'_\text{z} {\hat a}^\dag	& 0 			& 0			& 0 &\Delta_\frac{3}{2}&0\\
	-\sqrt{\frac{3}{2}}\gamma_-	{\hat a}	& -\sqrt{\frac{3}{2}}\gamma'_- {\hat a}^\dag		& 0 			& 0 				&0&0& \Delta_\frac{3}{2}\\   
     \end{pmatrix}.
  			 	\label{eqn:int2}
\end{align}
If $|\Delta_\frac{1}{2}| \approx |\Delta_\frac{3}{2}| \gg |\gamma|$ for all single phonon coupling rates $\gamma$, we can adiabatically eliminate the $P$ manifold to get an effective two-level atom. Such situation occurs for different species and a range of principal quantum numbers. For instance, taking $^{133}$Cs, $n=100$ for $\ket{s}$, $n=101$ for $\ket{s'}$ and $\omega_{\rm osc} = 2\pi \times 3$ GHz (the example considered in the main text) yields $\Delta_\frac{1}{2} = 2\pi \times 283$ MHz and $\Delta_\frac{3}{2}=2\pi \times 263$ MHz. In order to simplify the treatment, we thus replace the detunings in eq.(\ref{eqn:int2}) by $\Delta \approx \Delta_\frac{1}{2} \approx \Delta_\frac{3}{2}$. This also motivates the introduction of the effective transition frequency $\omega_{\rm a}'$ between the combined $P$ manifold and the $\ket{s'}$ state such that $\Delta = \omega_{\rm a}' - \omega_{\rm osc}$.

We are now in a position to apply the methods of degenerate perturbation theory \cite{shavitt:80} to find an effective Hamiltonian in the space spanned by $\{|s\rangle,|s'\rangle\}$. Defining the projector $\hat P = |s\rangle\langle s| + |s'\rangle \langle s'|$ and its complement $\hat Q = \mathds{1} - \hat P$, the Hamiltonian is partitioned into the block diagonal part $H_D = \hat P \hat H_\text{I} \hat P + \hat Q \hat H_\text{I} \hat Q$ and the off-diagonal perturbation $\hat V_\text{x} = \hat P \hat H_\text{I} \hat Q + \hat Q \hat H_\text{I} \hat P$.  We find a unitary transformation $\hat U = e^{\hat G}$, with $\hat G = -\hat G^\dagger$, such that $\hat H_\text{eff} = \hat U \hat H_\text{I} \hat U^\dagger$ is block diagonal, i.e. $\hat H_\text{eff} = \hat P \hat H_\text{eff} \hat P + \hat Q \hat H_\text{eff} \hat Q$ and $\hat{G} = \sum_{j = 0}^{\infty} \frac{1}{\Delta^j} G^{(j)}$.

The first non-zero contribution to the effective Hamiltonian is first order in $\frac{1}{\Delta}$ (second order in the expansion):

\begin{align}
\hat H_\text{eff} =  H_D + \frac{1}{2 \Delta} [G^{(1)},\hat V_x],\text{ with } G^{(1)} =  \begin{pmatrix}
   	0 				& 0					&-\gamma_- {\hat a}^\dag		& -\gamma_\text{z} {\hat a}^\dag	&\frac{1}{\sqrt 2}\gamma_- {\hat a}^\dag		& \sqrt 2 \gamma_\text{z} {\hat a}^\dag & -\sqrt{\frac{3}{2}}\gamma_+ {\hat a}^\dag  \\
	0				& 0					& -\gamma'_- {\hat a}		& -\gamma'_\text{z} {\hat a}& \frac{1}{\sqrt 2}\gamma'_- {\hat a}		& \sqrt 2 \gamma'_\text{z} {\hat a} & -\sqrt{\frac{3}{2}}\gamma'_+ {\hat a}\\
	-\gamma_+ {\hat a}		& -\gamma'_+ {\hat a}^\dag		& 0		& 0 				& 0 &0&0\\
	-\gamma_\text{z} {\hat a}	& -\gamma'_\text{z} {\hat a}^\dag	& 0 			& 0			& 0 &0&0\\
		\frac{1}{\sqrt 2}\gamma_+ {\hat a}		& \frac{1}{\sqrt 2}\gamma'_+ {\hat a}^\dag		& 0		& 0 				& 0 &0&0\\
	\sqrt 2 \gamma_\text{z} {\hat a}	& \sqrt 2 \gamma'_\text{z} {\hat a}^\dag	& 0 			& 0			& 0 &0&0\\
	-\sqrt{\frac{3}{2}}\gamma_-	{\hat a}	& -\sqrt{\frac{3}{2}}\gamma'_- {\hat a}^\dag		& 0 			& 0 				&0&0& 0\\   
     \end{pmatrix}
\end{align}
 
The resulting Hamiltonian in the space $\{|s\rangle,|s'\rangle\}$ is
\begin{align}
  \hat P \hat H_\text{eff} \hat P&= 
    \frac{3 \hbar}{\Delta}\begin{pmatrix}
   	\hat a^\dagger \hat a (\gamma_- \gamma_+ + \left.\gamma_z\right.^2)	
   	& (\hat a^\dagger)^2 \frac{1}{2}(\gamma'_- \gamma_+ + \gamma_- \gamma'_+ + 2 \gamma_z \gamma'_z) 	\\
	\hat a^2 \frac{1}{2}(\gamma'_- \gamma_+ + \gamma_- \gamma'_+ + 2 \gamma_z \gamma'_z) 				
	& \hat a^\dagger \hat a (\gamma'_- \gamma'_+ + \left.\gamma'_z\right.^2)					\\   
     \end{pmatrix}.
  			 	\label{eqn:int2eff}
\end{align}

The diagonal terms are the dispersive frequency shifts. A quasi-perfect two-photon-two-phonon resonance is achieved if $\frac{3\hbar n}{\Delta}\left[(\gamma_- \gamma_+ + \left.\gamma_z\right.^2) - (\gamma'_- \gamma'_+ + \left.\gamma'_z\right.^2)\right]$, with $n$ the phonon number, is negligibly small as compared to the off-diagonal terms in (\ref{eqn:int2eff}). For sufficiently small $n$ which is the situation of this article, and under the realistic assumption of $\mu_0 \approx \mu_0'$ the quasi-perfect resonance can be achieved and we thus consider only the off-diagonal terms of (\ref{eqn:int2}). The effective two-phonon coupling rate $\gamma_2$ is given by the off-diagonal terms

\begin{align}
\gamma_2(t) = \frac{3}{2 \Delta} \left( \gamma'_- \gamma_+ + \gamma_- \gamma'_+ + 2 \gamma_z \gamma'_z \right) = \left(\frac{Q z_\text{osc}}{ 4\pi\hbar\varepsilon_0 R^5}\right)^2\frac{\mu_0\mu'_0}{\Delta} \frac{1}{3}[R^2(R^2+3 Z^2)] \label{eqn:gamma2}
\end{align}
and the resulting interaction picture Hamiltonian reads

\begin{align}
  \hat H_\text{I,2}(t) \approx \hbar \gamma_2(t) |s\rangle\langle s'| (\hat a^\dagger)^2 + \hbar \gamma_2^*(t) |s'\rangle\langle s|\hat a^2 
\end{align}
The integrated coupling strength, for an atom taking a path $\mathbf{R}(t) = \{v t, 0, Z_0\}$ then becomes

\begin{align}
G_2 = \int_{-\infty}^\infty dt \gamma_2(t) = \left(\frac{Q z_\text{osc}}{4\pi\hbar\varepsilon_0}\right)^2\frac{\mu_0\mu'_0}{\Delta} \frac{21 \pi}{48 v Z_0^5} .
\end{align}

\subsection{Atomic dipole - oscillator quadrupole coupling}
\label{app:quadrupole}

In principle, the two-phonon resonance condition with interaction Hamiltonian similar to (\ref{eqn:Ham2}) can be achieved by exploiting the coupling between the atomic dipole and the oscillator quadrupole as we now show. The oscillator quadrupole corresponds to the $\hat z^2$ term in the expansion of $\hat E(\bf{r})$. Specifically, the $O(\hat z^2)$ term in (\ref{eqn:muE}) reads

\begin{align}
O(\hat z^2) = -\frac{Q \mu_0}{4\pi\varepsilon_0 R^7 } \frac{3}{2}\Big[
  				\hat M_{\text x} X(5 Z^2 - R^2)
  			  +\hat M_{\text y} Y(5 Z^2 - R^2)\nonumber
  			  + \hat M_{\text z} Z(5 Z^2 - 3 R^2)\Big]
  			 	\hat z^2 + O(\hat z^3) \label{eqn:Direct}
\end{align}

Under the two-phonon resonance condition $\omega_\text{osc} = \omega_\text{a}/2$, (\ref{eqn:Direct}) dominates the atom-oscillator interaction, and the resulting interaction picture Hamiltonian reads

\begin{align}
  \hat H_\text{I, quad}	&=  \frac{1}{2}\frac{Q \mu_0 z_\text{osc}^2}{4\pi\varepsilon_0 R^7}Z( 5 Z^2 - 3 R^2)\Big[				
  				|s\rangle\langle p| (\hat a^\dagger)^2    					   	 			
  					   	 + |p\rangle\langle s|\hat a^2
  				   			\Big] + F' \nonumber\\
  				   			&\approx \hbar \gamma_\text{2,quad}(t) |s\rangle\langle p| (\hat a^\dagger)^2 + \text{H. c.},
\end{align}
where  $F'$ contain only terms oscillating at $\omega_\text{osc}$ or higher frequency, which can be neglected through the rotating wave approximation, and $\gamma_\text{2,quad} =\frac{ Q \mu_0 z_\text{osc}^2}{4\pi\varepsilon_0 \hbar R^7} \frac{Z( 5 Z^2 - 3 R^2)}{2}$ is the two-phonon coupling strength.

For an atom trajectory $\mathbf{R}(t) = \{v t, 0, Z_0\}$  the integrated two-phonon coupling strength is  $G_\text{2,quad} = \int_{-\infty}^{\infty} \gamma_\text{2,quad} dt =  \frac{Q \mu_0 z_\text{osc}^2}{4 \pi \epsilon_0 \hbar} \frac{2}{3 v Z_0^3}$.

\section{Derivation of the master equation}
\label{app:master eq}

We start the derivation of the master equation for the single-phonon resonance by using (\ref{eq:U_n}) and the atomic initial state (\ref{eqn:init}) and find the state of the oscillator after the passage of a single atom. For brevity we will write the state before the $k$th atom has passed $\rho = \rho_\text{osc}^{(k-1)}\otimes \rho_\text{a}$. Expanding (\ref{eq:evolution}) yields

\begin{align}
&\rho_\text{osc}^{(k)} = \text{Tr}_\text{a}[\hat U_n \rho \hat U_n^\dagger] = \sum_{n,m = 0}^{\infty} \rho_{nm}\nonumber\\
&\Big[
  |\beta|^2 |n\rangle \cos(\Theta_n)\cos(\Theta_m) \langle m| 
	+  |\alpha|^2 |n-1\rangle \sin(\Theta_{n-1}) \sin(\Theta_{m-1}) \langle m-1|\nonumber\\
&- i\alpha\beta^* |n-1\rangle \sin(\Theta_{n-1})\cos(\Theta_m) \langle m|
	+  i\beta\alpha^* |n\rangle \cos(\Theta_n)\sin(\Theta_{m-1}) \langle m-1| \nonumber\\
&+ |\alpha|^2 |n\rangle \cos(\Theta_{n-1})\cos(\Theta_{m-1}) \langle m| 
	+  |\beta|^2 |n+1\rangle \sin(\Theta_n) \sin(\Theta_m) \langle m+1|\nonumber\\
&- i\beta\alpha^* |n+1\rangle \sin(\Theta_n)\cos(\Theta_{m-1}) \langle m|
	+  i\alpha\beta^* |n\rangle \cos(\Theta_{n-1})\sin(\Theta_m) \langle m+1| \Big],
\end{align}
where $\Theta_n = G \sqrt{n+1}$, and $\rho^{(k-1)}_\text{osc} = \sum_{n,m = 0}^{\infty} \rho_{nm} |n\rangle \langle m|$. In a similar fashion to the derivation in \cite{Englert:02} we transform the sum over $n$ and $m$ into an operator equation. Firstly, we can rewrite the bras and kets as  $|n-1\rangle = \frac{\hat a}{\sqrt{\hat a^\dagger \hat a}} |n\rangle$ and $|n+1\rangle = \frac{\hat a^\dagger}{\sqrt{\hat a \hat a^\dagger}} |n\rangle$. Secondly $n$ and $n+1$ are written as $\hat a^\dagger \hat a$ and $\hat a \hat a^\dagger$, resulting in the replacements

\begin{align} 
&|n\rangle\cos (G \sqrt n)\rightarrow\cos (G \sqrt{\hat a^\dagger \hat a})|n\rangle\\
&|n\rangle\cos (G \sqrt{n+1})\rightarrow\cos (G \sqrt{\hat a \hat a^\dagger})|n\rangle\\
&|n-1\rangle\sin (G \sqrt n)\rightarrow\frac{\hat a \sin (G \sqrt{\hat a^\dagger \hat a})}{\sqrt{\hat a^\dagger \hat a}}|n\rangle\\
&|n+1\rangle\sin (G \sqrt{n+1})\rightarrow\frac{\hat a^\dagger\sin (G \sqrt{\hat a \hat a^\dagger})}{\sqrt{\hat a \hat a^\dagger}}|n\rangle
\end{align}

This lets us replace $\sum_{nm}\rho_{nm}|n\rangle \langle m|$ with $\rho^{(k-1)}_\text{osc}$ giving

\begin{align}
\rho^{(k)}_\text{osc} =
&  |\beta|^2 \left[
	  \cos (G \sqrt{\hat a \hat a^\dagger})\rho^{(k-1)}_\text{osc}
	  	\cos (G \sqrt{\hat a \hat a^\dagger})
	+ \sin (G \sqrt{\hat a \hat a^\dagger})
			\frac{\hat a^\dagger}{\sqrt{\hat a \hat a^\dagger}}
						\rho^{(k-1)}_\text{osc} 
							\frac{\hat a}{\sqrt{\hat a \hat a^\dagger}}
								\sin (G \sqrt{\hat a \hat a^\dagger})
	\right]\nonumber\\
& + |\alpha|^2 \left[
  \cos (G \sqrt{\hat a^\dagger\hat a} )\rho^{(k-1)}_\text{osc}\cos (G \sqrt{ \hat a^\dagger\hat a})
+ \sin (G \sqrt{ \hat a^\dagger \hat a})\frac{\hat a}{\sqrt{\hat a^\dagger\hat a} } \rho^{(k-1)}_\text{osc} \frac{\hat a^\dagger}{\sqrt{\hat a^\dagger\hat a }}\sin (G \sqrt{\hat a^\dagger\hat a} )
\right]\nonumber\\
 &+  i\alpha\beta^* \left[
  \cos (G \sqrt{\hat a^\dagger\hat a} ) \rho^{(k-1)}_\text{osc}
\frac{\hat a}{\sqrt{\hat a \hat a^\dagger}}\sin (G \sqrt{\hat a \hat a^\dagger})
- \sin (G \sqrt{ \hat a^\dagger \hat a})\frac{\hat a}{\sqrt{\hat a^\dagger\hat a} }  \rho^{(k-1)}_\text{osc} 
\cos (G \sqrt{\hat a \hat a^\dagger})
\right]\nonumber\\
& + i\beta\alpha^* \left[
   \cos (G \sqrt{\hat a \hat a^\dagger})\rho^{(k-1)}_\text{osc}
  \frac{\hat a^\dagger}{\sqrt{\hat a^\dagger\hat a }}\sin (G \sqrt{\hat a^\dagger\hat a} )
- \sin (G \sqrt{\hat a \hat a^\dagger})\frac{\hat a^\dagger}{\sqrt{\hat a \hat a^\dagger}}\rho^{(k-1)}_\text{osc} 
\cos (G \sqrt{ \hat a^\dagger\hat a})
\right]
\end{align}

Note that, up until now, these equations remain exact. We are now interested in an approximation where $\langle n|G \hat a^\dagger \hat a|n\rangle \ll 1$ for all oscillator levels $n$ up to some maximum $n_\text{max}$ that we set as a truncation of the oscillator space. To second order  
 $\cos (G \sqrt{\hat a^\dagger \hat a})\approx 1-\frac{G^2}{2} \hat a^\dagger \hat a$,
$\cos (G \sqrt{\hat a \hat a^\dagger})\approx 1-\frac{G^2}{2} \hat a \hat a^\dagger $,
$\sin (G \sqrt{\hat a^\dagger \hat a})\frac{\hat a}{\sqrt{\hat a^\dagger \hat a}} \approx G \hat a$ and 
$\sin (G \sqrt{\hat a \hat a^\dagger})\frac{\hat a^\dagger}{\sqrt{\hat a \hat a^\dagger} } \approx G \hat a^\dagger$ leaving

\begin{align}
\rho^{(k)}_\text{osc} = &  |\beta|^2 \left(
\rho^{(k-1)}_\text{osc} + G^2 \left[\hat a^\dagger \rho^{(k-1)}_\text{osc} \hat a  - \frac{1}{2}( \hat a \hat a^\dagger \rho^{(k-1)}_\text{osc} + \rho^{(k-1)}_\text{osc} \hat a \hat a^\dagger ) \right]
\right)\nonumber\\
& + |\alpha|^2 \left(
\rho + G^2 \left[\hat a \rho^{(k-1)}_\text{osc} \hat a^\dagger  - \frac{1}{2}( \hat a^\dagger \hat a \rho^{(k-1)}_\text{osc} + \rho^{(k-1)}_\text{osc} \hat a^\dagger  \hat a) \right]
\right)\nonumber\\
 &+  i\alpha\beta^* G \left( \rho^{(k-1)}_\text{osc} \hat a - \hat a  \rho^{(k-1)}_\text{osc} \right)\nonumber\\
& + i\beta\alpha^* G \left(\rho^{(k-1)}_\text{osc}\hat a^\dagger-\hat a^\dagger\rho^{(k-1)}_\text{osc}  \right) + O(G^3)
\end{align}

We then can find our approximate master equation:

\begin{align}
\dot \rho^{(k)}_\text{osc} &\approx r\,( \rho^{(k)}_\text{osc} - \rho^{(k-1)}_\text{osc} ) \label{eq:master single supp}\\
&= r \left( \mathcal D[\alpha G \hat a](\rho_\text{osc}) + \mathcal  D[\beta G \hat a^\dagger](\rho_\text{osc}) -i G \left[ \alpha \beta^* \hat a + \beta \alpha^* \hat a^\dagger , \rho_\text{osc}\right]\right)
\end{align}

where we have used $|\alpha|^2 + |\beta|^2 = 1$, and for the last line the index $k$ has been suppressed, as none of the dynamics depend on it. The derivation of the master equation for the two-phonon resonance follows the same lines, with $\hat a(\hat a^\dagger)$ replaced by $\hat a^2((\hat a^\dagger)^2)$. 

The steady state of the evolution under (\ref{eq:master single supp}) is a displaced thermal state $\rho_{\rm osc} = D(A) \rho_{\rm th} D^\dag(A)$, where $\rho_{\rm th} = \sum_{n=0}^{\infty} \ket{n}\bra{n} \left(\frac{\bar n}{1+\bar n}\right)^n \frac{1}{1+\bar n}$ is the thermal state with average occupation number $\bar{n} = \frac{|\beta|^2}{|\alpha|^2-|\beta|^2}$, $D(A) = {\rm exp}\left[ A \hat{a}^\dag - A^* \hat{a}\right]$ is the coherent displacement operator and $A = i \frac{2 \alpha^* \beta }{G(|\alpha|^2 - |\beta|^2)}$ is the coherent shift amplitude. The solution is valid for values of $|\beta|^2$ below 0.5 as it becomes unstable for higher $\beta$ (negative $\bar n$).

\section{Squeezing angle}
\label{app:squeezing}

A system described by a Hamiltonian $\hat H =  \Omega (e^{-i \theta}\hat a^2 + e^{i\theta}(\hat a^\dagger)^2)$ evolves according to the operator 
$\hat S = \exp[-i \hat H t]$ 
$ = \exp[-i\Omega t (e^{-i\theta}\hat a^2 + e^{-i\theta}(\hat a^\dagger)^2)]$ 
$ = \exp[\Omega t (e^{-i(\theta+\pi/2)}\hat a^2 - e^{i(\theta+\pi/2)}(\hat a^\dagger)^2)]$. This yields the following operator relations

\begin{align}
\hat S^\dagger \hat a \hat S &= \hat a \cosh(\Omega t/2) - \hat a^\dagger e^{i(\theta+\pi/2)} \sinh(\Omega t/2) \label{eqn:squeeze1}\\
\hat S^\dagger \hat a^\dagger \hat S &= \hat a^\dagger \cosh(\Omega t/2) - \hat a e^{-i(\theta+\pi/2)} \sinh(\Omega t/2) \label{eqn:squeeze2}
\end{align}

We can now calculate the variance in the $\phi$ quadrature with a vacuum initial state $|0\rangle$, with $\hat \chi_\phi = \left(\hat a e^{-i\phi} + \hat a^\dagger e^{i\phi}\right)/\sqrt{2}$.

\begin{align}
\Delta \chi^2_\phi = \langle 0| \hat S^\dagger \hat \chi_\phi^2 \hat S |0\rangle - 
					\langle 0| \hat S^\dagger \hat \chi_\phi \hat S |0\rangle^2 \label{eqn:Var}
\end{align}

Using relations (\ref{eqn:squeeze1}) and (\ref{eqn:squeeze2}), (\ref{eqn:Var}) becomes

\begin{align}
\Delta \chi^2_\phi = &\frac{1}{2}\left(\cosh^2(\Omega t/2)+\sinh^2(\Omega t/2)\right) \nonumber\\
	&- \sinh(\Omega t/2)\cosh(\Omega t/2) \cos (2\phi - \theta -\pi/2) \nonumber\\
	= &\frac{1}{2}\big(\cosh(\Omega t) -\sinh(\Omega t) \cos (2\phi - \theta-\pi/2)\big)
\end{align}

For $\theta = 0$, as considered in the main text, $\Delta \chi^2_\phi$ is minimized for $\phi = \pi/4$.%

\end{widetext}

\end{document}